\documentstyle[preprint,aps,epsf,floats]{revtex}
\begin{document}
\tighten
\def\si{{}^1\kern-.14em S_0}
\def\siii{{}^3\kern-.14em S_1}
\def\diii{{}^3\kern-.14em D_1}
\def\pone{{}^3\kern-.14em P_1}
\def\pzero{{}^3\kern-.14em P_0}
\def\ptwo{{}^3\kern-.14em P_2}
\newcommand{\gsim}{\raisebox{-0.7ex}{$\stackrel{\textstyle >}{\sim}$ }}
\newcommand{\lsim}{\raisebox{-0.7ex}{$\stackrel{\textstyle <}{\sim}$ }}
\def\pislash{ {\pi\hskip-0.6em /} }
\def\pislashsmall{ {\pi\hskip-0.375em /} }
\def\pslash{p\hskip-0.45em /}
\def\nopi{ {\rm EFT}(\pislash) }
\def\Ltwo{ {^\pislashsmall \hskip -0.2em L_2 }}
\def\Lone{ {^\pislashsmall \hskip -0.2em L_1 }}
\def\CQuad{ {^\pislashsmall \hskip -0.2em C_{\cal Q} }}
\def\Czero{ {^\pislashsmall \hskip -0.2em C_{0}^{(\siii)} }}
\def\Czeromone{ {^\pislashsmall \hskip -0.2em C_{0,-1}^{(\siii)} }}
\def\Czerozero{ {^\pislashsmall \hskip -0.2em C_{0,0}^{(\siii)} }}
\def\Czeroone{ {^\pislashsmall \hskip -0.2em C_{0,1}^{(\siii)} }}
\def\Ctwo{ {^\pislashsmall \hskip -0.2em C_{2}^{(\siii)} }}
\def\Ctwomtwo{ {^\pislashsmall \hskip -0.2em C_{2,-2}^{(\siii)} }}
\def\Ctwomone{ {^\pislashsmall \hskip -0.2em C_{2,-1}^{(\siii)} }}
\def\Cfour{ {^\pislashsmall \hskip -0.2em C_{4}^{(\siii)} }}
\def\CSDzero{ {^\pislashsmall \hskip -0.2em C_0^{(sd)} }}
\def\CSDtwotwotwo{ {^\pislashsmall \hskip -0.2em C_{2,-2}^{(sd)} }}
\def\CSDzeromone{ {^\pislashsmall \hskip -0.2em C_{0,-1}^{(sd)} }}
\def\CSDzerozero{ {^\pislashsmall \hskip -0.2em C_{0,0}^{(sd)} }}
\def\CSDtwoone{ {^\pislashsmall \hskip -0.2em \tilde C_2^{(sd)} }}
\def\CSDtwotwo{ {^\pislashsmall \hskip -0.2em C_2^{(sd)} }}
\def\CSDzerotwo{ {^\pislashsmall \hskip -0.2em C_{0,0}^{(sd)} }}
\def\LX{ {^\pislashsmall \hskip -0.2em L_X }}
\def\CSDfour{ {^\pislashsmall \hskip -0.2em C_4^{(sd)} }}
\def\CSDfourt{ {^\pislashsmall \hskip -0.2em \tilde C_4^{(sd)} }}
\def\CSDfourtt{ {^\pislashsmall \hskip -0.2em {\tilde{\tilde C}}_4^{(sd)} }}
\def\etasd{\eta_{sd} }
\def\ZCzeromone{ 
{_z \hskip -0.4em {^\pislashsmall \hskip -0.2em C_{0,-1}^{(\siii)} }}}
\def\ZCzerozero{ 
{_z \hskip -0.4em {^\pislashsmall \hskip -0.2em C_{0,0}^{(\siii)} }}}
\def\ZCzeroone{ 
{_z \hskip -0.4em {^\pislashsmall \hskip -0.2em C_{0,1}^{(\siii)} }}}
\def\ZCtwomtwo{ 
{_z \hskip -0.4em {^\pislashsmall \hskip -0.2em C_{2,-2}^{(\siii)} }}}
\def\ZCtwomone{ 
{_z \hskip -0.4em {^\pislashsmall \hskip -0.2em C_{2,-1}^{(\siii)} }}}
\def\ZCfourmthree{ 
{_z \hskip -0.4em {^\pislashsmall \hskip -0.2em C_{4,-3}^{(\siii)} }}}
\def\rCzeromone{ 
{_\rho \hskip -0.4em {^\pislashsmall \hskip -0.2em C_{0,-1}^{(\siii)} }}}
\def\rCzerozero{ 
{_\rho \hskip -0.4em {^\pislashsmall \hskip -0.2em C_{0,0}^{(\siii)} }}}
\def\rCzeroone{ 
{_\rho \hskip -0.4em {^\pislashsmall \hskip -0.2em C_{0,1}^{(\siii)} }}}
\def\rCtwomtwo{ 
{_\rho \hskip -0.4em {^\pislashsmall \hskip -0.2em C_{2,-2}^{(\siii)} }}}
\def\rCtwomone{ 
{_\rho \hskip -0.4em {^\pislashsmall \hskip -0.2em C_{2,-1}^{(\siii)} }}}
\def\rCfourmthree{ 
{_\rho \hskip -0.4em {^\pislashsmall \hskip -0.2em C_{4,-3}^{(\siii)} }}}
\def\CPzero{ {^\pislashsmall \hskip -0.2em C^{(\pzero)}_2  }}
\def\CPone{ {^\pislashsmall \hskip -0.2em C^{(\pone)}_2  }}
\def\CPtwo{ {^\pislashsmall \hskip -0.2em C^{(\ptwo)}_2  }}

\def\Journal#1#2#3#4{{#1} {\bf #2}, #3 (#4)}

\def\JPA{\em J. Phys. A}
\def\NCA{\em Nuovo Cimento}
\def\NIM{\em Nucl. Instrum. Methods}
\def\NIMA{{\em Nucl. Instrum. Methods} A}
\def\NPB{{\em Nucl. Phys.} B}
\def\NPA{{\em Nucl. Phys.} A}
\def\NP{{\em Nucl. Phys.} }
\def\PLB{{\em Phys. Lett.} B}
\def\PRL{\em Phys. Rev. Lett.}
\def\PRD{{\em Phys. Rev.} D}
\def\PRC{{\em Phys. Rev.} C}
\def\PRA{{\em Phys. Rev.} A}
\def\PR{{\em Phys. Rev.} }
\def\ZPC{{\em Z. Phys.} C}
\def\SJP{{\em Sov. Phys. JETP}}
\def\SJNP{{\em Sov. J. Nucl. Phys.}}
\def\FBS{{\em Few Body Syst.}}
\def\IJMP{{\em Int. J. Mod. Phys.} A}
\def\UJP{{\em Ukr. J. of Phys.}}
\def\CJP{{\em Can. J. Phys.}}
\def\SCI{{\em Science} }
\def\AST{{\em Astrophys. Jour.} }

\preprint{\vbox{
\hbox{ NT@UW-99-42}
}}
\bigskip
\bigskip

\title{Improving the Convergence of $NN$  Effective Field Theory 
}
\author{Daniel R. Phillips$^a$, Gautam Rupak$^a$ and Martin J. Savage$^{a,b}$}
\address{$^a$ Department of Physics, University of Washington, \\
Seattle, WA 98195. }
\address{$^b$ Jefferson Lab., 12000 Jefferson Avenue, Newport News, \\
Virginia 23606.}

\maketitle

\begin{abstract} 
We study a low-energy effective field theory (EFT)
describing the $NN$ system in which all exchanged particles are
integrated out.  We show that fitting the residue of the ${}^3S_1$
amplitude at the deuteron pole, rather than the ${}^3S_1$ effective
range, dramatically improves the convergence of deuteron observables
in this theory. 
Reproducing the residue ensures that
the tail of the deuteron wave function, which is directly related to
$NN$ scattering data via analytic continuation, is correctly reproduced
in the EFT at next-to-leading order. 
The role of multi-nucleon-electroweak operators
which  produce deviations from effective-range theory
can then be explicitly separated
from the physics of the wave function tail. 
Such an operator contributes to the deuteron
quadrupole moment, $\mu_Q$,  at low order, 
indicating a sensitivity to short-distance physics.
This is
consistent with the failure of impulse-approximation calculations in
$NN$ potential models to reproduce $\mu_Q$. 
The convergence of $NN$
phase shifts in the EFT is unimpaired by the use of this new
expansion.  
\end{abstract}

\vskip 2in

\leftline{August 1999}
%
%
%
%
\vfill\eject


When the deuteron is probed at low momentum-transfers it is
appropriate to treat the nuclear bound state in a long-wavelength
approximation. In such an approximation no details of the
short-distance dynamics are observed. Consequently most observables
are dominated by the physics of the deuteron's wave function
tail. Ignoring, for the time being, the deuteron's D-state, two
numbers characterize this tail. 
One is the deuteron
binding energy, which determines the rate of exponential fall-off of
the wave function. 
The other is the asymptotic S-state normalization, 
$A_S$, which multiplies this exponential. 
In the limit as $r \rightarrow \infty$ the wave function becomes
\begin{equation} 
\psi({\bf r}) \quad \rightarrow \quad \frac{A_S}{\sqrt{4 \pi}} 
\frac{e^{-\gamma r}}{r} \, \equiv \, \sqrt{\frac{\gamma Z_d}{2 \pi}}
\frac{e^{-\gamma r}}{r}
\ \ \ .
\label{eq:psiER}
\end{equation}

Below we work in terms of the dimensionless quantity $Z_d$, rather than $A_S
\equiv \sqrt{2 \gamma Z_d}$. Neither $Z_d$ nor $A_S$ can be measured
directly in experiment, but their values can be determined by analytic
continuation of the $NN$ scattering amplitude into the bound-state
region. Parameterizing the ${}^3S_1$ phase shifts as~\cite{ERtheory}
\begin{eqnarray}
  |{\bf k}|\cot\delta_0 & = & -\gamma \ +\
  {1\over 2}\rho_d (|{\bf k}|^2+\gamma^2)\ +\
  w_2\ (|{\bf k}|^2+\gamma^2)^2\ +\ ...
\ \ \ ,
\label{eq:kcot}
\end{eqnarray}
with $\gamma^{-1} = 4.319\ {\rm fm}$,
$\rho_d = 1.765(4) \ {\rm fm}$, and
$w_2=0.389\ {\rm fm^3}$\cite{PSA,deSwart},
reproduces the phase-shift data in the
low-momentum region. 
Near the deuteron pole, the scattering amplitude becomes
\begin{eqnarray}
\lim_{k \rightarrow i \gamma}
{4\pi\over M_N} {1\over  |{\bf k}|\cot\delta_0 - i  |{\bf k}|}
\ = \ -{4\pi\over M_N} {1\over 1-\gamma\rho_d}\ {1\over i k + \gamma}
+ R(k)
\ \equiv \ 
 -{4\pi\over M_N}\ \frac{Z_d}{i k + \gamma} + R(k) \ \ \ .
\label{eq:kcotpole}
\end{eqnarray} 
Here the quantity $R(k)$ is regular
in the limit $k \rightarrow i \gamma$ and, importantly, is the only
place that higher-order terms in the expansion
of $|{\bf k}|\cot\delta_0$ enter this amplitude. It follows 
that the residue $Z_d$ is given by
\begin{equation} 
Z_d=\frac{1}{1 - \gamma \rho_d} 
\ \ \ .
\label{eq:Zddef}
\end{equation}
The measured values of $\gamma$ and $\rho_d$ quoted above give
$Z_d=1.690(3)$, which is also the value of $Z_d$ obtained using  
$A_S$ given in the 1993 Nijmegen phase-shift
analysis~\cite{PSA,deSwart}, as required.

In the 50
years since effective range theory was first formulated many
calculations have taken the ``effective-range-theory wave function"
given in eq.~(\ref{eq:psiER}) 
as the wave function for {\it all} $r$ and calculated
low-momentum processes involving the deuteron. As we will recapitulate
below, the results obtained by this procedure are often in good
agreement with both experimental data and modern $NN$ potential
models. This suggests that such quantities truly are insensitive to
details of the short-distance deuteron dynamics.

The central tenet of effective field theory (EFT) is that
low-momentum probes of any system should not reveal details of that
system's short-distance dynamics. Much effort has recently been invested
in calculating observables in the two-nucleon system using effective
field theory techniques based on this
tenet~\cite{Weinberg1}--\cite{Rhotalk}. An extreme, but efficacious,
EFT has been constructed for the $NN$ system by
considering only distance scales $r \ll 1/m_\pi$, i.e. the region
where the S-state deuteron wave function is given by
eq.~(\ref{eq:psiER}). In this EFT the low-momentum scales in the
problem are $\gamma$ and the typical nucleon momentum $k$.  
It is a non-relativistic theory of nucleons
interacting via contact interactions, which we denote by $\nopi$.
These contact interactions
represent the results of integrating the pions, and all other exchanged
degrees of freedom, out of the theory.
The leading contact
interaction is responsible for binding the deuteron, and it 
is enhanced over expectations based on naive
dimensional analysis, as are all other S-wave
contact interactions. Relativistic corrections can be included
perturbatively, and are very small. We shall not discuss them here,
the interested reader should consult Ref.~\cite{CRSa} for details. 
The Lagrangian describing the strong interactions in
$\nopi$ is
\begin{eqnarray}
{\cal L} & = &  
 - \Czero \left(N^T P_i N\right)^\dagger\left(N^T P_i N\right)
 -   \Ctwo  {1\over 2}
\left[(N^T P_i N)^\dagger
\left(N^T {\cal O}^{jj,i}_2 N\right) +  h.c.\right]
\ +\ \ldots \ \ \ ,
\label{eq:lagtwo}
\end{eqnarray}
where $P_i$ is the spin-isospin projector for the 
$\siii$ channel\cite{KSW,KSW2}
and the ellipses denote terms involving more spatial derivatives.
The two derivative operator, ${\cal O}^{xy,j}_2$, is given by 
\begin{eqnarray}
{\cal O}^{xy,j}_2 & = &
-{1\over 4}\left(
      \overleftarrow {\bf D}^x  \overleftarrow {\bf D}^y P^j
+ P^j \overrightarrow {\bf D}^x  \overrightarrow {\bf D}^y
- \overleftarrow {\bf D}^x P^j\overrightarrow {\bf D}^y
-\overleftarrow {\bf D}^y P^j\overrightarrow {\bf D}^x
\right)
\ \ \ .
\label{eq:twoder}
\end{eqnarray}

The coefficients $C_0$, $C_2$, $\ldots$ are to be fit to data. 
Until now
this has been done by reproducing the effective-range expansion 
(\ref{eq:kcot}) up to some
given order in the expansion parameter $Q \sim k, \gamma$. 
This is
straightforward, since the exact result for $|{\bf k}|\cot \delta$ in
$\nopi$, if relativity and S-D mixing are ignored,  is
\begin{equation}
  - |{\bf k}|\cot\delta_0 = 
        \frac{4 \pi}{M_N} \frac{1}{\sum_n C_{2n} |{\bf k}|^{2n}} + \mu
\ \ \ , 
\label{eq:kcotdsolved} 
\end{equation}
when dimensional regularization with power-law divergence subtraction
is employed\cite{KSW}.  Expanding each $C_{2n}$ in powers of $Q$
itself, i.e. writing: $C_{2n}=C_{2n,-n-1} + C_{2n,-n} + \ldots$, and
then expanding eq.~(\ref{eq:kcotdsolved}) in powers of $Q$ one can
match to eq.~(\ref{eq:kcot}) order-by-order in $Q$.  If $\rho_d$ and
$w_2$ are both taken to scale as $Q^0$, then at
next-to-next-to-leading order (N$^2$LO) in $Q$ this yields the
coefficients

\begin{eqnarray}
  \rCzeromone & = & -{4\pi\over M_N}{1\over  (\mu-\gamma)}
\ \ ,\ \ 
  \rCzerozero \ =\  {2\pi\over M_N}{ \gamma^2\rho_d\over (\mu-\gamma)^2}
\ \ ,\ \ 
  \rCzeroone \ =\  -{\pi\over M_N} {\gamma^4\rho_d^2\over (\mu-\gamma)^3}
\ \ ,\nonumber\\
  \rCtwomtwo & = &   {2\pi\over M_N}{ \rho_d\over (\mu-\gamma)^2}
\ \ ,\ \ 
  \rCtwomone \ =\  -{2\pi\over M_N}{ \gamma^2\rho_d^2 \over (\mu-\gamma)^3}
  \ \ , \ \ 
  \rCfourmthree \ =\  -\frac{\pi}{M_N} \frac{\rho_d^2}{(\mu - \gamma)^3} 
   \ \ , \label{eq:Cs} 
\end{eqnarray} 
at the renormalization scale $\mu$. The dominant interaction
$\rCzeromone$ must be iterated to all orders to generate the leading
order (LO) amplitude. All other interactions, including the
corrections to $C_0$, can be treated in perturbation theory. At
next-to-leading order (NLO) and above the coefficients are chosen to
reproduce $\rho_d$ exactly. Hence we refer to this choice as the
``$\rho$-parameterization.  With this choice of $C$'s, observables are
power-series expansions in the quantity $\gamma\rho_d$ order-by-order
in the $Q$ expansion. In particular, the expansion of $Z_d$ is
\begin{equation}
Z_d^{(Q)}\ =\ 1+\gamma\rho_d+(\gamma\rho_d)^2+(\gamma\rho_d)^3+ \ldots 
\ \ .
\label{eq:Zrho}
\end{equation}
Physically, the amplitude for any low-energy elastic reaction on the
deuteron will always include an overall factor of $Z_d$. Thus the
appearance of this series in a number of previous calculations in
EFT($\pislash$) should come as no surprise. Indeed, Park {\it et al.}
have had considerable success in reproducing experimental data by
pursuing a strategy of fitting $Z_d$ at NLO~\cite{Parka,Parkeft}.
In fact, such an approach is implicit in other work on EFT in nuclear 
physics, e.g. Refs.~\cite{Bira,GPLa,Epel}. 
Recently, Phillips and Cohen~\cite{PC99} and Rho~\cite{Rhotalk} have
stressed that it is better to fit $Z_d$ in the EFT
than $\rho_d$, since this ensures that the calculation correctly
reproduces the long-distance piece of the deuteron wave function. 
The
difference between fitting $\rho_d$ and $Z_d$ is higher-order in any
given EFT calculation, but, as we shall show here,
in EFT($\pislash$) fitting $Z_d$ markedly improves the convergence of
calculations for a variety of low-energy elastic processes on deuteron
targets. We will also discuss inelastic processes such as $n p \rightarrow d
\gamma$, and show that demanding that $Z_d$ is correctly reproduced
also improves the convergence of these calculations. Finally, we will
discuss the $NN$ phase shifts in the $Q$-expansion when this
alternative fitting procedure is employed.  We will show that, despite
abandoning the exact fit of $\rho_d$ in the $Q$-expansion of $k \cot
\delta$, the results found for the phase shift 
with this modified fitting procedure are no worse than those found
using the ``$\rho$-parameterization".

In this work we require that the position and residue of the deuteron
pole are reproduced exactly at NLO in the $Q$ expansion.  The
coefficients of the Lagrangian (\ref{eq:lagtwo}) determined by
this constraint are (again neglecting relativistic effects):

\begin{eqnarray}
  \ZCzeromone & = & -{4\pi\over M_N}{1\over  (\mu-\gamma)}
\ \ ,\ \ 
  \ZCzerozero \ =\  {2\pi\over M_N}{ \gamma (Z_d-1)\over (\mu-\gamma)^2}
\ \ ,\ \ 
  \ZCzeroone \ =\  
-{\pi\over M_N} {(Z_d-1)^2 (2\mu-\gamma)\gamma\over (\mu-\gamma)^3}
\ \ ,\nonumber\\
  \ZCtwomtwo & = &   {2\pi\over M_N}{Z_d-1\over \gamma (\mu-\gamma)^2}
\ \ ,\ \ 
  \ZCtwomone 
  \ =\  -{2\pi\over M_N}{(Z_d-1)^2 \mu  \over \gamma (\mu-\gamma)^3}
  \ \ , \ \ 
  \ZCfourmthree
  \ =\ -\frac{\pi}{M_N}\frac{(Z_d-1)^2}{\gamma^2 (\mu - \gamma)^3} \ \ .
\label{eq:ZCs} 
\end{eqnarray} 

We call this the ``$z$-parameterization''\footnote{Pronounced
``zed"-parameterization.}.  Here the quantity $Z_d-1$
is taken to be of order $Q$. This choice for the $C$'s is 
formally equivalent order-by-order
to that of eq.~(\ref{eq:Cs}), the difference between the two
always being of higher order in the $Q$-expansion. 
The advantage of this choice can be seen
in the $Q$-expansion of $Z_d$. Instead of the
convergence displayed in eq.~(\ref{eq:Zrho}) we now have, by explicit
construction:

\begin{equation} 
Z_d^{(Q)}=1 + (Z_d - 1) + 0 + 0 + 0 + \ldots \ \ .
\label{eq:Zz}
\end{equation}

Both parameterizations yield $Z_d^{(LO)}=1$. However, at NLO the two
series in eqs.~(\ref{eq:Zrho}) and (\ref{eq:Zz}) have very different
behavior. In the $\rho$-parameterization the first $\gamma\rho_d$
correction brings that expansion perturbatively close to the complete
sum of $Z_d$, with corrections of order $(\gamma\rho_d)^2$ and
higher. In contrast, in eq.~(\ref{eq:Zz}) the first $(Z_d-1)$
correction brings the series into exact agreement with the complete
sum, and all higher-order terms vanish.  This behavior translates
directly into the convergence of the $Q$-expansion for observables
involving the deuteron.


The easiest place to see this is in the deuteron charge form factor
$F_C$.  In the $\rho$-parameterization, neglecting relativistic
effects gives LO, NLO, and N$^2$LO contributions to the charge form
factor which are~\cite{CRSa},

\begin{eqnarray}
  F_C^{\rho (0)} ( |{\bf q}|) & = & {4\gamma\over |{\bf q}|} \tan^{-1}\left(
    { |{\bf q}|\over 4\gamma}\right)
\ \ \ ,\ \ \  F_C^{\rho (1)} ( |{\bf q}|) \ =\  - \gamma\rho_d \left[ 1 - 
{4\gamma\over |{\bf q}|} \tan^{-1}\left( { |{\bf q}|\over 4\gamma}\right)
\right] \ \ ,
\nonumber\\
  F_C^{\rho (2)} ( |{\bf q}|) & = & 
- \gamma^2\rho_d^2 \left[ 1 - 
{4\gamma\over |{\bf q}|} \tan^{-1}\left( { |{\bf q}|\over 4\gamma}\right)
\right]
\ -\ 
{1\over 6} r_N^2 \  |{\bf q}|^2\ 
{4\gamma\over |{\bf q}|} \tan^{-1}\left(
    { |{\bf q}|\over 4\gamma}\right) \ \ \ ,
\label{eq:EFTcffrho}
\end{eqnarray} 
where $|{\bf q}|$ is the momentum transfer, and
$r_N=0.790\pm 0.012~{\rm fm}$ 
is the isoscalar charge radius of the nucleon. In
contrast, the $z$-parameterization yields LO, NLO and N$^2$LO
contributions to the form factor:

\begin{eqnarray}
  F_C^{z (0)} ( |{\bf q}|) & = & {4\gamma\over |{\bf q}|} \tan^{-1}\left(
    { |{\bf q}|\over 4\gamma}\right)
\ \ \ ,\ \ \ 
F_C^{z (1)} ( |{\bf q}|) \ =\ 
 -(Z_d-1) \left(1 - {4\gamma\over |{\bf q}|} \tan^{-1}\left(
    { |{\bf q}|\over 4\gamma}\right)\right) \ \ , 
\nonumber\\
  F_C^{z (2)} ( |{\bf q}|) & = & 
-{1\over 6} r_N^2 \  |{\bf q}|^2\ 
{4\gamma\over |{\bf q}|} \tan^{-1}\left(
    { |{\bf q}|\over 4\gamma}\right) \ \ \ .
\label{eq:EFTcffz}
\end{eqnarray} 

Comparing the expressions for $F_C$ in the $\rho$ and $z$ 
parameterizations it is clear that demanding that $Z_d$ be reproduced exactly
at NLO means that the {\it only} correction at N$^2$LO is due to the 
finite size of the nucleon. There are no corrections resulting from
``wave function" effects at any order beyond NLO. 

This can be clearly understood if we assume the wave function given in
eq.~(\ref{eq:psiER}), is correct for all $r$, calculate $F_C$ via:

\begin{equation} 
F_C(|{\bf q}|)=\int \, d^3r \, e^{i {\bf q} \cdot
{\bf r}/2} |\psi({\bf r})|^2, 
\end{equation} 
and also add an additional (constant) piece to $F_C$ which ensures
that $F_C(0)=1$. When this is done we reproduce
eq.~(\ref{eq:EFTcffz}), but with $r_N=0$.  In other words, at NLO the
$z$-parameterization exactly reproduces the results obtained in
effective range theory calculations of a deuteron containing
point-like nucleons. However, EFT also allows for systematic
incorporation of the effects of finite-nucleon size, via terms such as
those proportional to $r_N^2$ above, in a way that maintains the
locality of the field theory.

Furthermore, one order beyond that calculated here, i.e. at N$^3$LO, a
four-nucleon-one-photon interaction contributes to $F_C$ and 
shifts the deuteron charge radius from its impulse approximation
value. This contribution to $F_C$ is beyond the scope of effective
range theory. 
In contrast, the charge radius extracted from
eq.~(\ref{eq:EFTcffz})  at N$^2$LO is
\begin{eqnarray}
 \langle r_d^2\rangle^{\rm EFT} & = & r_N^2
  \ +\ 
 {Z_d\over 8\gamma^2} \ =\ 4.566~{\rm fm}^2
 \ \ \ \ .
\label{eq:EFTcr} \end{eqnarray} We estimate the error from the next-order 
corrections (including the counterterm) to be $\sim3\%$. In
Table~\ref{table-staticprops} we compare this, and other static
properties of the deuteron, to those obtained in the potential models
Nijm93, Reid93~\cite{nijmegen}, and OBEPQ~\cite{Bonnreport}
(energy-independent version), in impulse approximation, and also to
the experimental value. We see that in the case of
$\langle r_d^2 \rangle$ the EFT calculation agrees
with data and potential model calculations within 1--2 \%.
Note that in comparing results we have assumed a common contribution
of $r_N^2=(0.79 \,\,\rm{fm})^2$ to $\langle r_d^2 \rangle$.

\begin{table}[h]
\begin{center}
\begin{tabular}{|c|c|c|c|}
  \hline 
Calculation & $\langle r_d^2 \rangle$ (fm$^2$) & $\mu_Q$ (fm$^2$) 
  & $\langle r_Q^2 \rangle$ (fm$^4$) \\ \hline \hline 
   EFT (LO)   &   2.332      &  0.335        &    0.70   \\
   EFT (NLO)  &   4.566      &  0.286 (fit)  &    1.40   \\
   OBEPQ      &   4.499      &  0.274        &    1.27   \\
   Nijm93     &   4.489      &  0.271        &    1.22   \\
   Reid93     &   4.501      &  0.270        &    1.22   \\
   Experiment & 4.493(23) & 0.2859(3) & 
\end{tabular} 
\end{center}
\caption{\label{table-staticprops} 
Results for the square of the deuteron charge radius, $\langle r_d^2 \rangle$,
the deuteron quadrupole moment, $\mu_Q$, and the square of the quadrupole 
``radius", $\langle r_Q^2 \rangle$, at LO and NLO, compared with 
impulse-approximation potential-model
calculations, and the experimental results.
The experimental value of $\langle
r_d^2 \rangle$ is taken from Ref.~\protect\cite{Wong}.
See also the more recent Refs.~\protect\cite{buch,FMS}.}
\end{table}


In Table~\ref{table-staticprops} we have also displayed results for
the deuteron's quadrupole moment, $\mu_Q$, and for 
the quadrupole ``radius", $\langle r_Q^2 \rangle$. We now spend some
time discussing the results obtained in EFT($\pislash$) for the
deuteron's quadrupole form factor. This quantity clearly demonstrates
the limitations of effective range theory. 

Naively one might think that it is impossible for the deuteron to
acquire a quadrupole moment in this EFT, but such
is not the case. The deuteron becomes non-spherical due to the
appearance in the theory of contact operators with two derivatives
which mix the S and D-states. In Ref.~\cite{CRSb} it was shown that
the coefficient of this operator could be fit to the asymptotic D-to-S
ratio, $\eta_{sd}$. The value of $\eta_{sd}$ is known to high
precision from analytic continuation of the S-D mixing parameter,
$\overline{\varepsilon}_1$: $\eta_{sd}=0.02544 \pm 0.0002$~\cite{deSwart}. Fitting
$\eta_{sd}$ means that the tail of the D-wave piece of the deuteron
wave function will be correct. 
The leading order result for $F_Q$ is
then just the same as that obtained assuming asymptotic forms
for the deuteron radial wave functions, working to leading order in
$\eta_{sd}$, and setting $Z_d=1$
\begin{equation}
\frac{1}{M_d^2}  F_Q^{z (0)} ( |{\bf q}|)  =  
- {3 \etasd \over 2\sqrt{2} \gamma |{\bf q}|^3}
\left[ \ 
4 |{\bf q}| \gamma -\left(3 |{\bf q}|^2+16\gamma^2\right) 
\tan^{-1}\left({|{\bf q}|\over 4\gamma}\right)
\right] \ \ .
\label{eq:QFF0}
\end{equation}
At this order the deuteron quadrupole moment is 
$\mu_Q^{(LO)}=\etasd/(\sqrt{2}\gamma^2)=0.335~{\rm fm}^2$,
determined from the $|{\bf q}|\rightarrow 0$ 
limit of eq.~(\ref{eq:QFF0}) (see also Ref.~\cite{Bl88}).

In the $z$-parameterization at NLO for $F_Q$ corrections proportional
to $Z_d - 1$ appear.  Also at NLO there is a contribution from a
four-nucleon-one-quadrupole-photon local operator whose coefficient is
not related by gauge invariance to the $NN$ scattering phase shift
data~\cite{CRSa}.  This coefficient must be determined from the
deuteron quadrupole moment.  An overall shift in the value of $F_Q$,
which we denote by $\delta \mu_Q$, results from the inclusion of this
operator.  The total NLO contribution to $F_Q$ is
\begin{equation}
\frac{1}{M_d^2} F_Q^{z (1)} ( |{\bf q}|) \ =\ 
\delta\mu_{\cal Q} + (Z_d - 1) 
\left[\frac{1}{M_d^2} F_Q^{z (0)} ( |{\bf q}|)  - \mu_Q^{(LO)}\right] \ \ .
\label{eq:QFF1}
\end{equation}
The numerical value of $\delta \mu_Q$ required to reproduce the
measured value of the deuteron quadrupole moment is $\delta
\mu_Q=-0.0492$ fm$^2$.  The EFT analyses of
Refs.~\cite{CRSa,CRSb} suggest that a counterterm of this size is
expected from renormalization group arguments and naturalness. As it
is the contribution of a two-body quadrupole charge operator to $F_Q$
it simply does not appear in effective range theory.  In
Refs.~\cite{CRSa,PC99} it was argued that the absence of such a term could
be responsible for ``modern'' potential models' persistent
underprediction of $\mu_Q$ by $\sim 5$ \% in impulse
approximation~\cite{nijmegen,AV18,Ma96}. The preeminent role played by
this counterterm, as compared to others in the $NN$ effective field
theory, is because $\mu_Q$ is more sensitive to short-distance physics
than many other deuteron observables.

As in the charge form factor, the only correction to this result at
N$^2$LO comes from the finite-size of the nucleon.
\begin{equation}
\frac{1}{M_d^2} F_Q^{z (2)} (|{\bf q}|) \ =\
-{1\over 6} \ r_N^2\ |{\bf q}|^2 \ 
\frac{1}{M_d^2} F_Q^{z (0)}(|{\bf q}|)
 \ \ \ ,
\label{eq:QFF2}
\end{equation}
All the nuclear effects 
have already been taken care of by the demand that $Z_d$ be 
replicated exactly.

\begin{figure}[t]
\centerline{{\epsfxsize=3.5in \epsfbox{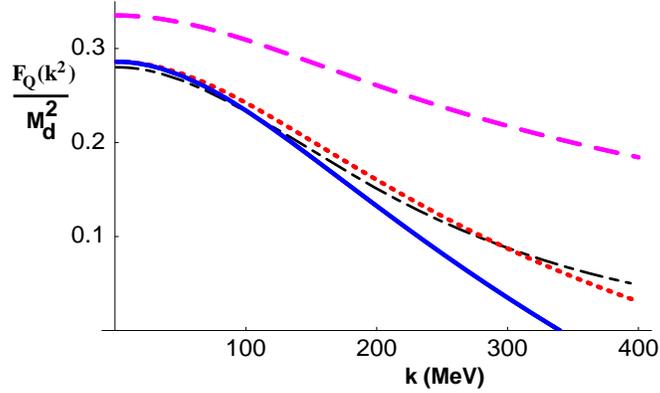}} } 
\noindent
\caption{\it The deuteron quadrupole form factor.  The dashed, dotted
and solid curves correspond to the LO, NLO and N$^2$LO contributions
in the $z$-parameterization.  The dot-dashed curve corresponds to a
calculation with the Bonn-B potential\protect\cite{Ma89} in the formulation of 
\protect\cite{AGA} . } 
\label{fig:QFF} 
\vskip .2in 
\end{figure} 
The form factor computed at LO, NLO and N$^2$LO is shown in
fig.~\ref{fig:QFF}, together with a potential-model calculation
which used the Bonn-B potential~\cite{Ma89} in the formulation of~\cite{AGA}.
The difference between $F_Q(|{\bf q}|=0)$ for the two higher-order
$\nopi$ calculations and the potential model calculation shown in
fig.~\ref{fig:QFF}, occurs because the calculation using the Bonn-B
potential does not produce the correct deuteron quadrupole moment.

We can also easily compute the deuteron quadrupole radius, defined by
\begin{equation} 
\langle r_Q^2 \rangle \equiv \left.-6
\frac{d F_Q}{d q^2} \right|_{q^2=0} \ \ \ .
\end{equation} 
At N$^2$LO there is
no counterterm contribution to $\langle r_Q^2 \rangle$, and
so we expect that it will be less sensitive to short-distance physics
than $\mu_Q$. Specifically:
\begin{equation} 
\langle r_Q^2 \rangle  =   {9\ \etasd\over 80
\sqrt{2} \gamma^4} \left( Z_d + {80\over 9} r_N^2 \gamma^2\right)
\ \ =\ \ 1.40~{\rm fm}^4 \ \ \ . \label{eq:rQ} 
\end{equation} 
At next order a contribution proportional to $\eta_{sd}^2$ enters, and
a two-body counterterm also appears. The error in our result for
$\langle r_Q^2 \rangle$ is dominated by the counterterm, which
produces a relative error $(\gamma/m_\pi)^5 1/\eta_{sd} \sim 15$
\%. This is roughly the discrepancy between eq.~(\ref{eq:rQ}) and the
results obtained from potential models with similar values of
$\eta_{sd}$.

Similar results to those shown here for $F_C$ and $F_Q$ can also be
obtained for the deuteron magnetic form factor, $F_M$.  It too has good
convergence properties in the $z$-parameterization of $\nopi$.


We now leave the form factors for elastic electron-deuteron scattering
and instead turn our attention to a quantity which is very much dominated
by the long-distance physics of the deuteron. The electric polarizability
of the deuteron has been computed in EFT($\pislash$) up to N$^3$LO in the
$\rho$-parameterization~\cite{CRSa,CRSb,CSbb}:

\begin{eqnarray}
\alpha_{E0}^{(\rho)}
     & = & 
{\alpha M_N\over 32\gamma^4}
    \left[ 1 \ \ +\ \  \gamma \rho_d\ \ +\ \  \gamma^2 \rho_d^2
    \ \ +\ \  \gamma^3 \rho_d^3
      \ \ +\ \ {2\gamma^2\over 3 M_N^2}
       \ \ +\ \ {M_N\gamma^3\over 3\pi} D_P
      \right]
\nonumber\\
& = & \qquad 0.377\ +\ 0.153\ +\ 0.062\ +\ 0.022\ \ +\ 0.0006\ \ -\ 0.0036\
\nonumber\\
& = & 0.616~{\rm fm}^3
\ \ \ ,
\label{eq:eftpolrho}
\end{eqnarray}
where $D_P=\CPzero + 2 \  \CPone + {20\over 3}\  \CPtwo 
\ =\  -1.51~{\rm fm}^3$ is the combination
of P-wave coefficients that contributes to the E1 amplitude in 
$np\rightarrow d\gamma$.
The dominant corrections here always
arise from the perturbative expansion of $Z_d$.
The relativistic correction, 
formally N$^2$LO in the momentum expansion,
gives  a contribution which numerically is of much higher order.
From now on we will ignore it.
The uncertainty in this calculation in the
$\rho$-parameterization is then set by the omission of  
$\gamma^4 \rho_d^4$ and higher. 
This constitutes an uncertainty of
$\sim 2.5\%$ due to the expansion of $Z_d$, and
leads to  $\Delta\alpha_{E0}^{(\rho)}\sim\pm 0.015$ fm$^3$.

In contrast, the $z$-parameterization gives
\begin{eqnarray}
\alpha_{E0}^{(z)}
     & = & 
{\alpha M_N\over 32\gamma^4}
    \left[ 1 \ \ +\ \  (Z_d-1) \ \ +\ \ 0 \ \ +\ \ {M_N\gamma^3\over 3\pi} D_P
      \right]
\nonumber\\
& = & \qquad 0.3770 \ \ +\ \ 0.2605 \ \ +\ \ 0 \ \ -\ \ 0.0036\
\ \ =\ \  0.6339~{\rm fm}^3
\ \ \ ,
\label{eq:eftpolz}
\end{eqnarray}
rapid convergence indeed. The insertion of the P-wave operators
also picks up factors of $Z_d$ when 
computed to higher orders and therefore we provocatively write
\begin{eqnarray}
\alpha_{E0}^{(z)}
     & = & 
{\alpha M_N\over 32\gamma^4}\ Z_d\ 
    \left[ 1\ \ +\  {M_N\gamma^3\over 3\pi} D_P\ +\ ...
      \right] \nonumber\\ & = & 0.6314~{\rm fm}^3 \ \ \ ,
\label{eq:eftpolzZ} 
\end{eqnarray}
where the ellipses denote higher-order terms, which 
include a N$^4$LO
contribution from $S-D$ mixing proportional to
$\eta_{sd}^2$. The first contribution from photon-four-nucleon
vertices which are not constrained by $NN$ phase shifts or gauge
invariance also occurs at N$^4$LO. We expect that these
two effects combined may shift the value of $\alpha_{E0}$ by
$\pm 0.0015$ fm$^3$.

This result is commensurate with the good agreement between ``modern''
potential model calculations which give $\alpha_{E0}=0.6328\pm 0.0017\
{\rm fm}^3$~\cite{FFa,FPa}.  The deuteron electric polarizability is very
precisely predicted once the $NN$ phase shifts are properly
described. EFT thus provides a very accurate
description of this quantity, as long as one demands that the tail of the
deuteron wave function be well-reproduced at low order in the EFT
expansion. This will be generally true for any low-energy elastic process
on the deuteron. Examples of such processes already computed in the
$\rho$-parameterization include Compton scattering~\cite{CGSSpol,Ccompt},
and the reaction $\nu d \rightarrow \nu d$~\cite{BCnu}.


We now turn our attention to reactions where the deuteron is only
present in the final or initial state. In particular, we wish
to see if the $z$-parameterization improves the convergence of
these reactions, in which only one of the asymptotic states involves
a deuteron.

The radiative capture process, $np\rightarrow d\gamma$, has been studied
in great detail in both $\nopi$\cite{CRSa,CRSb,CSbb} and the theory with
pions\cite{SSWst}. For incident cold or thermal neutrons, the rate for
this process is dominated by M1 capture from the $\si$ channel via the
isovector magnetic moment of the nucleon. The amplitudes for capture from
other channels are strongly suppressed in this kinematic
regime. Forthcoming experimental data from Grenoble has stimulated
interest in the calculation of the isoscalar M1 and E2 matrix elements in
$np \rightarrow d \gamma$ using different formulations of effective field
theory~\cite{PMRsupp,CRSb}.  In Ref.~\cite{CRSb} these amplitudes were
computed in the $\rho$-parameterization of $\nopi$. We now wish to examine
them in the $z$-parameterization.  The isoscalar M1 amplitude has been
computed up to NLO, and at that order it is the same in both the $\rho$
and $z$ parameterizations, due to the vanishing contribution of the
one-body operator. However, the isoscalar E2 amplitude is affected by the
choice of parameterization. Up to N$^2$LO the amplitude $\tilde X_{E2_s}$,
as defined in \cite{CRSb}, becomes
\begin{eqnarray} 
\tilde X_{E2_s} & = &  { \delta\mu_{\cal Q}\ \gamma^2\over4\sqrt{2}}
- {\etasd\over 10} \left( 1 + {3\over 8}(Z_d-1) -{7\over 32}(
Z_d-1)^2\right) + {3\over 40} E_1^{(4)}\
\gamma^4 \ \ \ ,
\label{eq:E2mat} 
\end{eqnarray} 
when the $z$-parameterization is used. 
Here $E_1^{(4)}$ is found by fitting the shape of the S-D mixing
parameter $\overline{\varepsilon}_1$ 
and turns out to be $E_1^{(4)} = -2.880~{\rm
fm}^4$~\cite{CRSb}.

Numerically, this leads to a series for the ratio of the isoscalar E2
matrix element to the experimental value of the dominant isovector M1
matrix element:
\begin{eqnarray}
\frac{\tilde{X}_{E2_{S}}}{\tilde{X}_{M1_{V}}} & = &   
- \left[\  1.565\ +\ 0.693\ +\ 0.209 \ \right]
\times 10^{-4}
\ =\ 
-2.47\times 10^{-4}
\ \ \ .
\label{eq:E2num}
\end{eqnarray}
While the central value is essentially unchanged from that found in
Ref.~\cite{CRSb} we now estimate the uncertainty in this ratio to
be $\sim 6\%$, much smaller than the $\sim 15\%$ quoted there.  The
series in eq.~(\ref{eq:E2num}) appears to be converging faster than the
analogous expression obtained using the
$\rho$-parameterization\cite{CRSb}. The reasons for this are not
entirely clear, since the naive expansion parameter, $Z_d-1=0.690$, is
actually {\it larger} than $\gamma \rho_d=0.408$. However, it seems
that, in general, an expansion of $\sqrt{1 + Z_d - 1}$ in powers of
$Z_d - 1$ converges faster than one of $(1 - \gamma \rho_d)^{-1/2}$
in powers of $\gamma\rho_d$.
It will be interesting to see how far the 
$z$-parameterization improves the convergence of EFT calculations of
the astrophysically-important reaction $pp \rightarrow d e^+
\nu_e$. Kong and Ravndal have pointed out that the
main obstacle to agreement between EFT and potential model
calculations of this reaction is the EFT's failure to exactly
reproduce the overall factor of $\sqrt{Z_d}$ which appears in the
astrophysical $S$-factor~\cite{Kong}.  

\begin{figure}[t]
\centerline{{\epsfxsize=3.5in \epsfbox{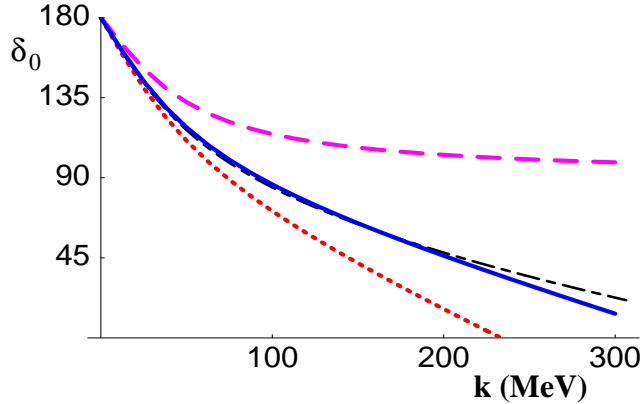}} }
\noindent
\caption{\it The $\siii$ phase shift as a function of nucleon 
momentum $|{\bf k}|$.
The dashed, dotted and solid curves correspond to the 
LO, NLO and N$^2$LO contributions in the $z$-parameterization.
The dot-dashed curve corresponds to the 
Nijmegen partial wave analysis\protect\cite{PSA}.}
\label{fig:ZDel}
\vskip .2in
\end{figure}

Finally, we turn our attention to $NN$ scattering. In this case we
merely wish to reassure ourselves that the $z$-parameterization does no
{\it worse} than the $\rho$-parameterization. We examine the $\siii$
phase shift $\overline{\delta}_0$ up to N$^2$LO. Taking the expression
for the coefficients (\ref{eq:ZCs}) and computing the amplitude, and
hence the phase shift, up to a given order in $Q$, we
find, successively:
\begin{eqnarray} 
\overline{\delta}_0 = \pi - \tan^{-1}\left({{\bf
k}\over \gamma}\right) \ -\ (Z_d-1) {{\bf k}\over 2 \gamma} \ +\
(Z_d-1)^2{{\bf k}\over 4 \gamma} \ \ , \label{eq:ZDel} 
\end{eqnarray}
where the first two pieces are order $Q^0$, the second is order $Q$, and
the third order $Q^2$. Relativistic corrections are again omitted. This
phase shift is plotted in fig.~\ref{fig:ZDel}.  At N$^2$LO it agrees very
well with the Nijmegen partial wave analysis~\cite{PSA}. The agreement is
certainly as good as that obtained in the $\rho$-expansion~\cite{CRSa}. It
is very encouraging that even for momenta which are large in the context
of this EFT, i.e. $k \sim m_\pi$, the convergence of
the $Q$-expansion for $NN$ scattering is good if the $z$-parameterization
is used. Given this result, and that observables involving the deuteron
appear to converge significantly faster, it is clear that the
$z$-parameterization of the coefficients in the Lagrange density
(\ref{eq:lagtwo}) is better than the $\rho$-parameterization that has been
used up until now. Its use may also improve the convergence of
three-nucleon observables such as the quartet phase
shifts~\cite{threebod,BG99}.

%

$\nopi$ is the most general theory in the two-nucleon sector consistent
with the symmetries of the strong and electroweak interactions in the very
low-energy regime. As we explained above, the only change implemented
here, vis \'a vis previous work on this theory, is to ensure that the tail
of the deuteron wave function, in both the S and D-states, is correctly
reproduced at low order in the EFT when elastic
processes are calculated. That this improves the convergence of the EFT is
not surprising. By choosing slightly different constraints to determine
the coefficients in the Lagrange density we have been able to write
observables in terms of the normalization constants $Z_d$ and $\eta_{sd}$,
rigorously defined in terms of the residue of the scattering amplitude at
the deuteron pole. If the coefficients are fit in this way then the full
impulse approximation result in effective range theory is obtained for
elastic processes at NLO in the EFT---at least up to
leading order in $\eta_{sd}$.  NLO $\nopi$ calculations
of elastic processes get the long-distance piece of the nuclear dynamics
correct. However, $\nopi$ calculations then go beyond effective range
theory, in that they systematically include two-body currents which
account for the short-distance physics not included in the asymptotic wave
functions. The advantage of using the EFT is that
these contributions, which are not constrained by $NN$ phase shift data,
have sizes set by the underlying short-distance physics. 
The best example of this is the
four-nucleon-one-quadrupole-photon operator which contributes at
NLO to the quadrupole moment.

\vskip 0.5in

GR and MJS thank Jiunn-Wei Chen and David Kaplan for useful
discussions. DRP is grateful to Tom Cohen for many conversations on
the tail of the wave function in EFT. This work is supported in part
by the U.S. Dept. of Energy under Grant No. DE-FG03-97ER4014.

\end{document}